\documentclass[aps,prb,twocolumn,longbibliography,noeprint]{revtex4-2}

\usepackage{graphicx,color,dcolumn,bm,amsfonts,amsmath,amssymb}
\usepackage{mathptmx,longtable,float}
\usepackage{titlesec,booktabs,eqnarray}
\usepackage{soul} 

\begin{document}

\title{Effect of Co Substitution on Ferrimagnetic Heusler compound Mn$_3$Ga }

\author{Quynh Anh T. Nguyen$^1$}
\author{Thi H. Ho$^1$}
\author{Myung-Hwa Jung$^2$}
\author{Sonny H. Rhim$^1$}
\email{sonny@ulsan.ac.kr}

\affiliation{$^1$Department of Physics and Energy Harvest Storage Research Center, 
University of Ulsan, Ulsan 44610, Republic of Korea\\
$^2$Department of Physics, Sogang University, Seoul, 04107, Republic of Korea}
\date{\today}

\begin{abstract}
Effect of Co substitution on Mn$_3$Ga is investigated using first-principles study
for structural and magnetic properties.
Without Co, ferrimagnetic Heusler compound Mn$_3$Ga is in tetragonal phase.
With Co substitution, 
depending on Co concentration ($x$) Mn$_3$Ga prefers
tetragonal (cubic) phase when $x\leq 0.5$ ($x\geq 0.5$).
Ferrimagnetism is robust regardless of $x$ in both phases.
While magnetic moments of two Mn do not vary significantly with $x$, 
Co magnetic moment in two phases exhibit different behaviors, 
leading to distinct features in total magnetic moment (${\rm M_{tot}}$).
When $x\leq 0.5$, in tetragonal phase, 
Co magnetic moment is vanishingly small, resulting in a decrease of ${\rm M_{tot}}$ with $x$.
In contrast, when $x \geq 0.5$, in cubic phase, 
Co magnetic moment is roughly 1$\mu_B$, which is responsible for an increase of ${\rm M_{tot}}$.
Electronic structure is analyzed with partial density of states for various $x$.
To elucidate the counterintuitively small Co moment,
the magnetic exchange interaction is investigated where exchange coefficient between Co and Mn is much smaller in $x\leq 0.5$ case than $x\geq0.5$ one.
\end{abstract}
\maketitle
  
\section{Introduction}
\label{sec:int}

For decades, Heusler compounds with plethora of material classes have drawn
continuously renewed attentions \cite{manna2018heusler,wollmann2017annuheusler}.
Rich physics has been platforms for  various research.
Numerous functionalities and properties, such as superconductivity, magnetism, multiferroicity, half-metallicity, topological insulators, and so forth \cite{graf2011:prog.ssc, kurtulus2005electronic,sargolzaei2006coyz,ishida1995search},
have attracted for applications in many areas.
In particular, in the context of rapid progress in magnetic devices and spintronics,
tetragonal Heusler compounds have been intensively and extensively explored.
Properties of tetragonal Heusler compounds,
perpendicular magnetic anisotropy (PMA) and low saturation magnetization ($M_S$) \cite{gasi2013exchange,faleev2017heusler,kundu2018first}, are regarded advantageous 
in applications with low switching current ($I_S$) \cite{Berger:STT,slonczewski96:jmmm,winterlik2012design}.

Mn$_3$Ga belonging to Heusler family in tetragonal structure,  possesses 
aforementioned magnetic properties: 
strong PMA around 0.89 MJ/m$^{-3}$, 
high Curie temperature ($T_C$) of 740 K \cite{kren1970neutron,balke2007mn}, 
low $M_S$ around 200 emu/cm$^3$, 
and approximately 58\% of spin polarization \cite{kurt2011high}.
Low $M_S$ stems from the intrinsic ferrimagnetism of antiparallel alignment of two Mn moments with unequal magnitudes.
Efforts  have been devoted to enhance the magnetic properties of Mn$_3$Ga.
One is to replace Mn by other transition metal (TM) \cite{sahoo2016compensated} 
where V-shape magnetization with respect to TM content ($x$) is quite distinct.
Upon substitution of nonmagnetic TM, ${\rm M_{tot}}$ decreases with $x$ 
as nonmagnetic TM replaces one Mn, moment is reduced but the antiparallel moment with the other Mn is retained.
${\rm M_{tot}}=0$ is achieved for certain $x$
above which 
the survived antiparallel moment dominates hence ${\rm M_{tot}}$ increases.

The V-shape of magnetization is also prominent with magnetic TM substitution
such as Fe, Co, and Ni \cite{felser2013tetragonal,nayak2015design}.
However, ${\rm M_{tot}}$ nearly vanishes above $x=0.5$ and is not compensated. 
Interestingly, Co substitution accompanies a structural transition from tetragonal to cubic
while the substitution by Fe and Ni retains the tetragonal structure.
Moreover, the magnetic behavior of magnetic TM in two structural phases is different. 
Magnetic moments in tetragonal phase are smaller than the bulk counterpart 
but those in cubic phase are comparable to the bulk ones.
While the substitution of Fe results in hard magnet and that of Ni exhibits the shape memory phenomena,
the property of Co substitution depends on the Co content.
It has been reported earlier that Co-poor ($0<x<0.5)$ in tetragonal alloy 
is well suited for spin-transfer torque (STT)
whereas Co-rich $(0.5<x<1)$ favors cubic phase with half-metallicity. 
As STT and half-metallicity are favorable components in spintronic application,
further study would shed light on uncovering Mn$_{3-x}$Co$_x$Ga.


In this paper, 
employing {\em ab initio} calculations, the effect of Co substitution on Mn$_{3-x}$Co$_x$Ga is investigated.
Magnetic property of Mn$_{3-x}$Co$_x$Ga is illustrated along with total as well as atomic resolved magnetic moments.
Analysis  follows to reveal the electronic structure of Mn$_{3-x}$Co$_x$Ga.
To elucidate the contrasting behavior of Co magnetism in two phases, tetragonal and cubic,
the exchange coefficients are discussed in the framework of the magnetic force theorem using the Heisenberg model.

\section{Structure and computational Methods}
\label{sec:comp}

\begin{figure}[t]
    \centering
    \includegraphics[width=0.75\columnwidth]{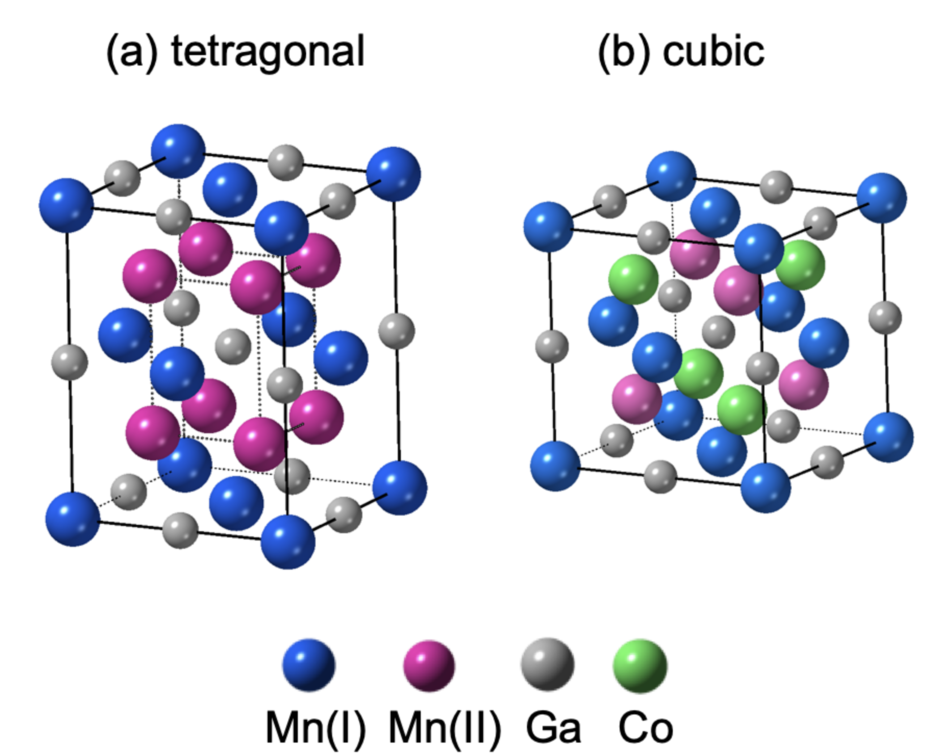}
    \caption{Crystal structure of (a) tetragonal Mn$_3$Ga and (b) cubic Mn$_2$CoGa. 
    Blue, red, green and grey spheres denote Mn(I), Mn(II), Co and Ga atoms, respectively.}
    \label{fig:pic1}
\end{figure}

Tetragonal ($x=0$) and cubic phases ($x=1$) are presented in Fig.~\ref{fig:pic1}, 
which correspond to $D0_{22}$ and $L_{21}$ structure, respectively.
Both tetragonal and cubic phase have two inequivalent Mn atoms distinguished by Mn(I) and Mn(II).
Space group of tetragonal phase is $I4/mmm$ (No. 139) and that of cubic phase is ${Fm\bar{3}m}$ (No. 225).
Wyckoff positions of both phases are listed in Table~\ref{tab:1}.

\begin{table}[b]
\centering
\caption{
Wyckoff positions of constituent atoms of tetragonal and cubic phases, whose space groups are 
$I4/mmm ~(No. 139)$ and $Fm\bar{3}m~(No. 25)$, respectively.
Co takes either $2c$ or $2d$ site.
}
\begin{ruledtabular}
  \begin{tabular}{ccc}
                       &  Tetragonal & Cubic \\
\hline
Mn(I)  & $2b$ & $4a$ \\
Mn(II) & $4d$ & $4c$ \\
Ga     & $2a$ & $4b$\\
Co     & 2c/2d&  $4d$  
\end{tabular}
\end{ruledtabular}
\label{tab:1}
\end{table}

First-principles calculations are performed using Vienna {\em Ab initio} Simulation Package (VASP) \cite{kresse1993ab,kresse1996efficient} with projector augmented-wave basis \cite{blochl1994projector}.
The exchange-correlation potential is treated using generalized gradient approximation (GGA) 
with Perdew, Burke, and Ernzerhof (PBE) parametrization \cite{perdew1992ja,perdew1996generalized}.
Cutoff energy of 500 eV is employed for wave function expansion.
For Brillouin zone integration, ${15\times15\times7}$ and ${15\times15\times11}$ {\em k} meshes are used for tetragonal and cubic phases, respectively, where $\sqrt{2}/2\times\sqrt{2}/2\times 1$ unit cell is adopted for the primitive cell.

From lattice constant optimization, 
we obtain $a = 3.78$ \AA\ ($c/a=1.88$) for tetragonal phase and $a$ = 4.12 \AA\ for cubic phase, respectively, 
which are used for further calculations 
 The exchange coefficients are calculated using the Heisenberg model 
 in the framework of the magnetic force theorem (MFT) \cite{liechtenstein1987local}.
To do so, additional self-consistent calculations are performed using OpenMX \cite{ozaki2003variationally},
where atomic basis sets for Mn, Ga, and Co are ${s3p2d1}$, ${s2p2d1}$, and H-$s3p2d1$, respectively
with energy cutoff of 300 Ry and cutoff radii of 10.0$\AA$ (Mn and Co) and 9.0$\AA$ (Ga), respectively.
Then, $Jx$ package is employed \cite{yoon2018reliability,yoon2020jx,han2004electronic}  
to extract the exchange coefficients ($J$) in the framework of MFT.

\section{Results and Discussion}
\label{sec:results}
\subsection{Magnetic Moments}
\label{sec:III-A}
\begin{figure}[t]
    \centering
    \includegraphics[width=\columnwidth]{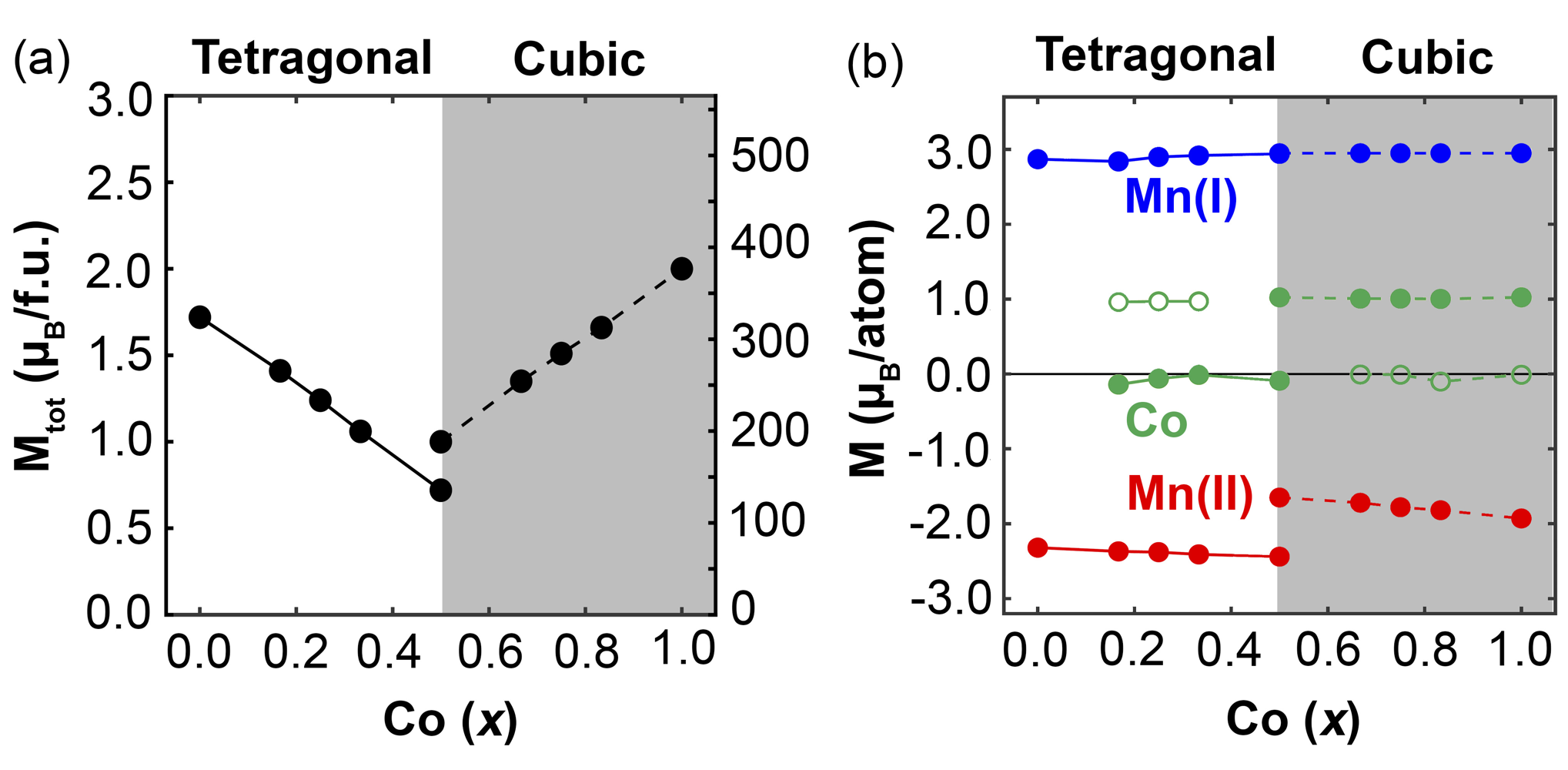}
    \caption{(a) Magnetic moment (M$_{tot}$) per unit cell with respect to $x$,
    where white and grey area are tetragonal and cubic phases, respectively. 
    Right axis is in emu/cm$^3$ converted from $\mu_B$ on the left axis.
    (b) Atom-resolved magnetic moment: blue, red, and green circles denote Mn(I), Mn(II), and Co, respectively.
    Open (solid) circles represent unstable (stable) phase.}

    \label{fig:pic2}
\end{figure}


Mn$_3$Ga is either tetragonal or cubic depending on Co concentration $x$.
When $x<0.5$, tetragonal phase with $c/a=1.88$ is favored whereas when $x>0.5$ cubic phase is preferred.
As such, $x=0.5$ is a phase boundary of two phaes \cite{SMTCE}.

Total magnetic moment (${\rm M_{tot}}$) 
as function of $x$ is plotted in Fig.~\ref{fig:pic2}, where
each moment of individual Co, Mn(I), and Mn(II) is also shown. 
For all $x$, Mn$_{3-x}$Co$_x$Ga is ferrimagnetic:
Moments of Mn(I) and Mn(II) are antiparallel with different magnitudes.
Without Co, i.e. $x=0$, Mn(I) and Mn(II) have moments of 2.87 and -2.32 $\mu_B$, respectively,
exhibiting indeed ferrimagnetism.
Total magnetic moment (${\rm M_{tot}}$) is 1.72 $\mu_B$ per formula unit, 
consistent with previous works \cite{yun2012strong,balke2007mn},
which corresponds to magnetization of $314$ emu/cm$^3$.
As shown in Fig.~\ref{fig:pic2}(a), ${\rm M_{tot}}$ decreases (increases) with $x$ in tetragonal (cubic) phase.
Notably, at  the phase boundary when $x=0.5$, 
${\rm M_{tot}}$ of cubic phase is larger by 0.4 $\mu_B$ than tetragonal phase.
This difference is due to different Co magnetism in both phases.
Acknowledging ferromagnetism of Co, Co moment in tetragonal phase is quite small \cite{SMLDAU},
which is against to the conventional wisdom.
This is discussed later in Sec.\ref{sec:III-B} and \ref{sec:III-C}.

Magnetic moments of individual atoms are shown in Fig.~\ref{fig:pic2}(b) \cite{SMGa}.
For all $x$, Mn(I) and Mn(II) have almost constant moments with
2.90 and -2.32 $\mu_B$, respectively.
However, Co moment in tetragonal phase nearly vanishes while in cubic phase is around 1.0 $\mu_B$.
Consequently,
${\rm M_{tot}}$ decreases with $x$ in tetragonal phase: 
${\rm M_{tot}}=1.72$ $\mu_B$ for $x=0$ drops to 0.72 $\mu_B$ for $x=0.5$.
On the other hand, in cubic phase, more Co substitution results in increase of ${\rm M_{tot}}$ from 1.00 to 2.00 $\mu_B$.
The vanishingly small Co magnetic moment in tetragonal phase is also reported in previous work 
which has been also reported in other theoretical work \cite{wollmann2015schemefor}.
As mentioned earlier, this counterintuitive feature will be visited in \ref{sec:III-C}.

\begin{figure}[t]
    \centering
    \includegraphics[width=\columnwidth]{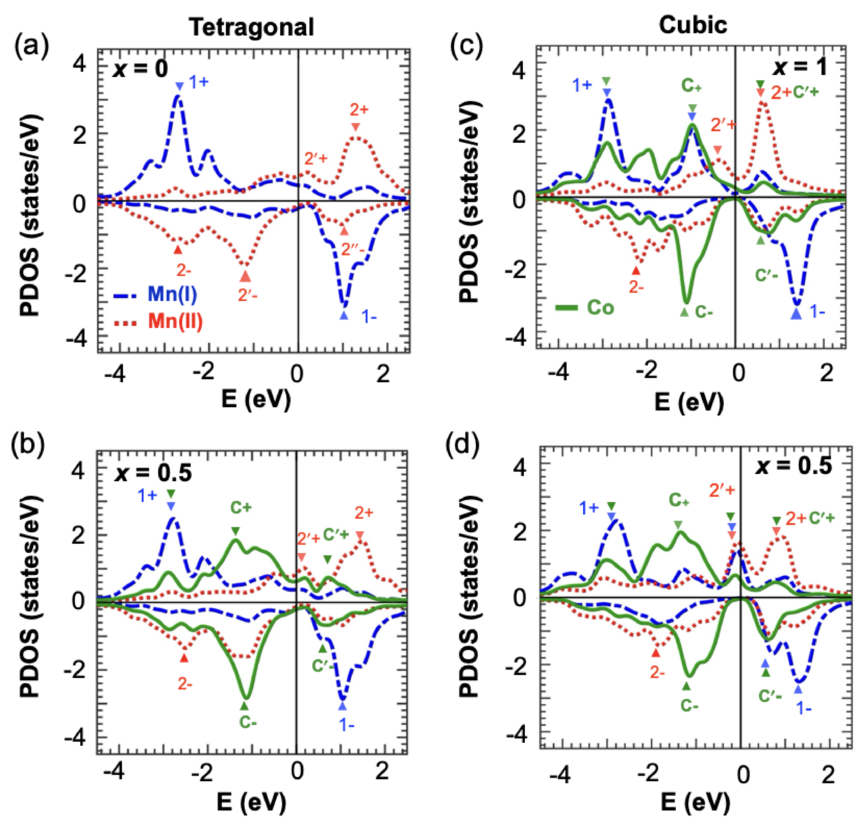}
    \caption{ Partial density of states (PDOS) of $d$ orbitals.
    Left and right panels for tetragonal and cubic phases, respectively.
    Tetragonal phase with (a) $x=0$ and (b) $x=0.5$.    
    Cubic phase with (c) $x=1.0$ and (d) $x=0.5$.
    Lower panel is comparison between tetragonal and cubic phases with $x=0.5$.
    Blue, red, and green lines denote $d$ orbitals of Mn(I), Mn(II), and Co, respectively.
    Peaks are emphasized with triangles. Fermi energy ($E_F$) is set to zero.
    }
    \label{fig:pic3}
\end{figure}

\subsection{Electronic Structure}
\label{sec:III-B}
Partial density of states (PDOS) is presented in Fig.~\ref{fig:pic3}.
Left (right) upper panel is tetragonal (cubic) phase for $x=0$ ($x=1$).
Lower panel is for $x=0.5$ of tetragonal and cubic phases, respectively.
Important peaks are labelled with triangle marks:
1, 2, and $C$ stand for Mn(I), Mn(II), and Co, respectively; 
+ (-) for the majority (minority) spin state, respectively.
Different peaks in the same spin state of particular atom are distinguished by primes.
PDOS's of $x=1/3$ and $2/3$ are shown in Fig. S3,
where the vertical line serves to manifest the overall trend upon $x$.
The shift of $E_F$ by Co substitution is evidence as Co has two more electrons than Mn\cite{SMMOM}.

In tetragonal phase with $x=0$ [Fig.~\ref{fig:pic3}(a)], Mn(I) and Mn(II) exhibit distinct features. 
Mn(I) has two main peaks: $1$+ ($1$-) in occupied (empty) states 
around -2.7 eV (+1.0 eV) in the majority (minority) spin state.
On the other hand, Mn(II) has rather broader peak with smaller height than Mn(I):
2+ around +1.3 eV in the majority spin state and 
2-, 2$'$-, and 2$''$- around -2.5, -1.2 eV, and +1.0 eV in the minority spin state, respectively.

For $x=0.5$ in tetragonal phase [Fig.~\ref{fig:pic3}(b)], 
Mn(I) and Mn(II) PDOS do not change much with respect to $x=0$ but shifted by +0.20 eV due to Co.
[See vertical line in Fig.~S3].
On the other hand, several peaks emerge in Co PDOS.
As Co replaces Mn(II), most Co peaks overlap with Mn(II):
$C$- coincides with 2$'$- and small hump just above $E_F$ overlap with 2$'$+.
Nonetheless, the overlap with Mn(I) in the majority spin state
around -2.8 eV (marked with double triangles) is evident.
Broad $C$+ appear between -1.4 and 0.4 eV 
owing to hybridization with Ga $p$ state [See Fig.~S2]. 
In particular, there are small humps around +0.2 and +0.6 eV, 
where the former, just mentioned earlier, overlaps with 2$'$+ but the latter has no such feature.
As the sum of occupied PDOS of two spin states is almost equal, Co magnetic moment nearly vanishes,
which is prominent in tetragonal phase.

Now we switch to cubic phase with $x=1$ [Fig.~\ref{fig:pic3}(c)].
Compared to $x=0$ of tetragonal phase, 1+ changes little but 1- shifts by +0.3 eV.
However, changes in Mn(II) and Co are apparent.
In the minority spin state, 2- shifts by 0.2 eV and 2$'$- is replaced by $C$-. 
Overlaps between 2$''$- (not labeled) and $C'$- are prominent around $E_F$+0.8 eV. 
On the other hand, in the majority spin state, $C$+ appears almost around the same place as $C$-.
Overlaps between $C'$+ and 2+ around $E_F$+0.6 eV are much closer to $E_F$ than $x=0$.
Notably, 2$'$+ is occupied in contrast to $x=0$. 

Cubic phase with $x=0.5$ is shown in Fig.~\ref{fig:pic3}(d).
Here, we discuss changes with respect to $x=0.5$ of tetragonal phase and $x=1$ of cubic phase.
Hence in this paragraphs $x=0.5$ ($x=1.0$) refers to that in tetragonal (cubic) phase.
Mn(I) is little affected: 1+ and 1- change negligibly with respect to $x=1$
whereas 1- shifts by +0.3 eV with respect to $x=0.5$.
On the other hand, Mn(II) is quite affected with  negative (positive) shifts with respect to $x=0.5$ ($x=1$).
More specifically, compared to $x=1$, 
peaks of 2+ and 2$'$+ shift by +0.30 and +0.20 eV, respectively; that of 2- by +0.35 eV.
Compared to $x=0.5$,
shifts of 2+ and 2$'$+ are -0.60 and -0.40 eV, respectively; 2- by -0.50 eV. 
In particular, 2$'$+ accompanies occupation change. 
For Co PDOS,
C+ and C- shift by -0.4 and -0.1 eV, respectively, with respect to $x=1$, 
and 0 and -0.1 eV with respect to $x=0.5$.
Positive shift of Mn(II) with respect to $x=1$ is well expected, as mentioned earlier, 
as Co has more electrons than Mn.

\begin{figure}[t]
    \centering
    \includegraphics[width=0.75\columnwidth]{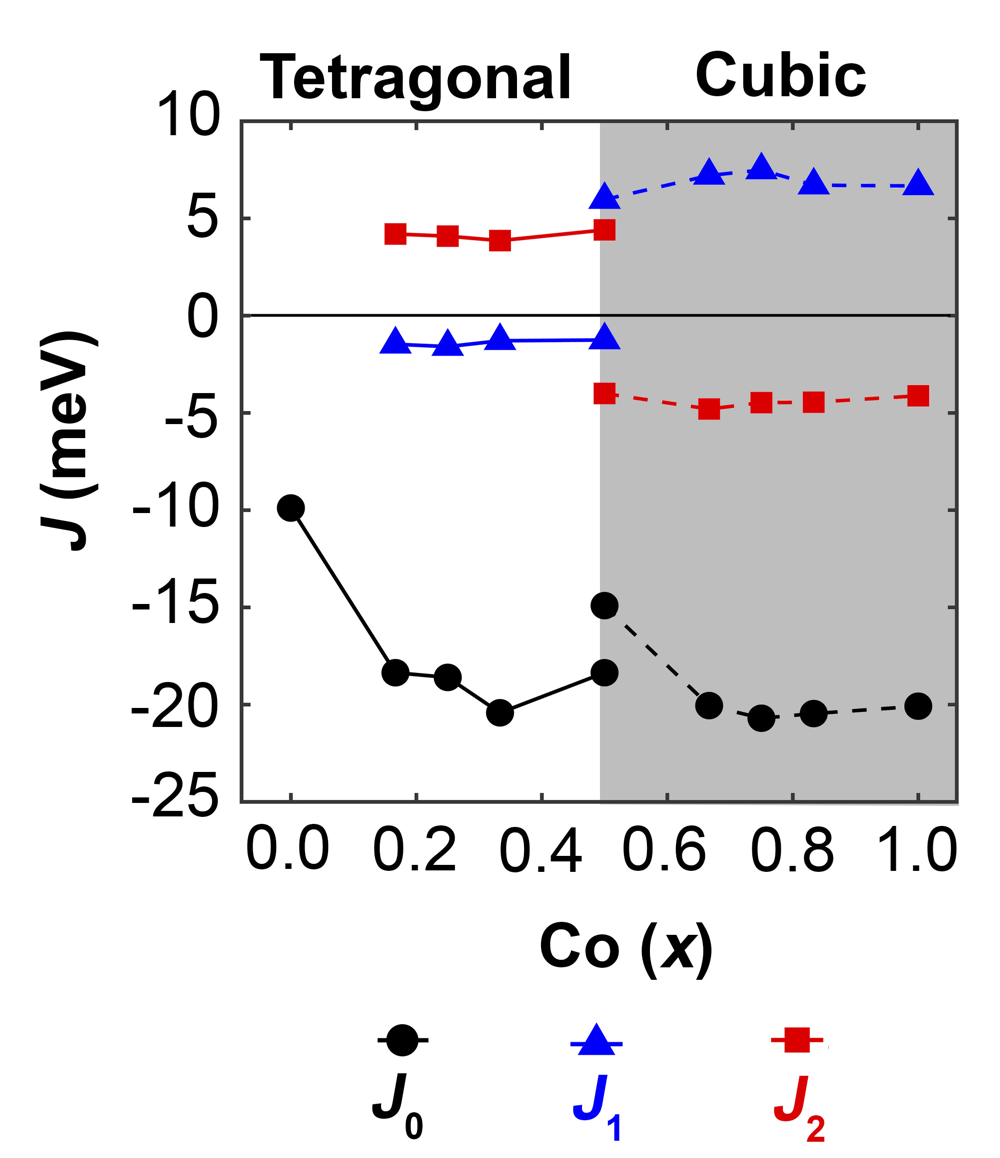}
    \caption{Calculated exchange coefficients ($J$) for nearest neighbor interactions for various $x$ ($0\leq x\leq$ 1). $J_0$, $J_1$, and $J_2$,  represent the exchange coefficients of Mn(I)-Mn(II), Co-Mn(I), Co-Mn(II) in black, blue, and red, respectively.
    Tetragonal and cubic regions are divided by white and grey regions.
    Solid and dashed lines for tetragonal and cubic phases, respectively.
    }
    \label{fig:pic4}
\end{figure}
\subsection{Magnetic Exchange Interaction}
\label{sec:III-C}
So far PDOS analysis is presented.
While Co magnetic moment in cubic phase ($x \geq 0.5$) is around 1 $\mu_B$, 
that in tetragonal phase $x \leq 0.5$ nearly vanishes, 
which is against the conventional wisdom.
In order to clarify this puzzling Co magnetism, the exchange coefficients, $J$, are discussed,
employing the Heisenberg model, $\hat{H}=-\sum_{ij}{J_{ij}{{S_i}}\cdot{S_i}}$,
where $J_{ij}$, the exchange coefficients, with positive (negative) sign implies parallel (antiparallel) magnetic coupling.
As briefly mentioned in \ref{sec:comp}, $J'$s are calculated in the framework of MFT \cite{liechtenstein1987local,yoon2018reliability,han2004electronic,ozaki2003variationally}. 
Fig.\ref{fig:pic4} shows calculations of the exchange coefficients for three cases, Mn(I)-Mn(II), Co-Mn(I), and Co-Mn(II), 
denoted as $J_0$, $J_1$, and $J_2$, respectively.

$J_0$, the exchange coefficients for two Mn types, is negative for all cases.
This indicates the antiparallel magnetism between two Mn sites 
confirming the ferrimagnetism.
Moreover, $|J_0|$ is larger than $|J_1|$ or $|J_2|$ roughly by an order of magnitude.
In tetragonal phase ($x \leq 0.5$), $J_0=-9.89$ meV for $x=0$ 
whose magnitude increases with $x$ ranging 18 -- 21 meV.
In cubic phase ($x\geq 0.5$), $J_0=-14.9$ meV for $x=0.5$,
whose magnitude also increases with $x$ up to 20 meV. 
From this,  addition of Co introduces stronger tendency of the antiparallel magnetism between Mn(I) and Mn(II).

The exchange coefficients between Co-Mn(I) and Co-Mn(II), namely $J_1$ and $J_2$, exhibit opposite sign behavior
where magnitudes in cubic phase are slightly larger.
In contrast to $J_0$, both $|J_1|$ and $|J_2|$ are little influenced by Co.
In tetragonal phase, $|J_1|$ and $|J_2|$ are approximately 1 and 5 meV, respectively, while
in cubic phase$|J_1|$ and $|J_2|$ are approximately 5 and 7 meV, respectively.   
The opposite signs of $J_1$ and $J_2$ is due to the antiparallel magnetism between Mn(I) and Mn(II).
From $|J_2|>|J_1|$, the antiparallel coupling between Co-Mn(II) is stronger than Co-Mn(I). 
Moreover,  $J_1$ and $J_2$ change signs between tetragonal and cubic phases,
henceforth Co shows the opposite sign of magnetic moment in both phases.
From the analysis of the exchange coefficients,
Mn(I) and Mn(II) exhibit rather rigid magnetic moments. 
Moreover, the exchange coefficients with Co in tetragonal phase are weaker than cubic phase. 
As a result, Co moments in tetragonal phase is smaller than cubic one. 

\section{Conclusion}\label{sec:concl}
In summary, we have investigated the effect of Co substitution on Mn$_3$Ga
for structural and magnetic properties.
The ferrimagnetic feature is robust regardless of Co concentration ($x$).
When $x \leq 0.5$, tetragonal phase is favored over cubic and vice versa, 
where the magnetic behavior of each phase is different.
In tetragonal phase, ${\rm M_{tot}}$ decreases with $x$
while in cubic phase ${\rm M_{tot}}$ increases with $x$.
Electronic structure upon Co substitution is investigated with PDOS, 
whose peak structures are analyzed.
Except Co, magnetic moments of individual atom change little with $x$.
Co magnetic moment in tetragonal phase nearly vanishes in contrast to 1$\mu_B$ of cubic phase.
To get more insight on this contrasting Co moment in two phases,
the exchange coefficients are estimated using the Heisenberg model.
Indeed, the antiparallel magnetism, more precisely, 
the ferrimagnetism between Mn(I)-Mn(II) with different magnitudes is well manifested.
On the other hand, the exchange coefficients of Co-Mn(I) and Co-Mn(II) exhibit opposite signs in tetragonal and cubic phase.

\begin{acknowledgments}
We are grateful to Soon Cheol Hong for fruitful discussions and comments.
This work was supported by the National Research Foundation (NRF) of Korea (2019R1I1A3A01059880, 2020R1A2C3008044).
\end{acknowledgments}

%
\end{document}